# Anisotropic spin-orbit torque generation in epitaxial SrIrO₃ by symmetry design

T. Nan[a,1], T. J. Anderson[a,1], J. Gibbons[b], K. Hwang[c], N. Campbell[d], H. Zhou[e], Y. Q. Dong[e], G. Y. Kim[f], N. Reynolds[b], X. J. Wang[g], N. X. Sun[g], S. Y. Choi[f], M. S. Rzchowski[d], Yong Baek Kim[c,h,i], D. C. Ralph[b,j] and C. B. Eom[a,2]

[a]Department of Materials Science and Engineering, University of Wisconsin-Madison, Madison, Wisconsin 53706, USA; [b]Laboratory of Atomic and Solid State Physics, Cornell University, Ithaca, New York 14853, USA; [c]Department of Physics and Centre for Quantum Materials, University of Toronto, Toronto, Ontario M5S 1A7, Canada; [d]Department of Physics, University of Wisconsin-Madison, Madison, Wisconsin 53706, USA; [e]Advanced Photon Source, Argonne National Laboratory, Argonne, Illinois 60439, USA; [f]Department of Materials Science and Engineering, POSTECH, Pohang 37673, Korea; [g]Department of Electrical and Computer Engineering, Northeastern University, Boston, Massachusetts 02115, USA; [h]Canadian Institute for Advanced Research/Quantum Materials Program, Toronto, Ontario M5G 1Z8, Canada; [i]School of Physics, Korea Institute for Advanced Study, Seoul 130-722, Korea; and [j]Kavli Institute at Cornell for Nanoscale Science, Ithaca, New York 14853, USA

**Abstract:**

  **Spin-orbit coupling (SOC), the interaction between the electron spin and the orbital angular momentum, can unlock rich phenomena at interfaces, in particular interconverting spin and charge currents. Conventional heavy metals have been extensively explored due to their strong SOC of conduction electrons. However, spin-orbit effects in classes of materials such as epitaxial 5d-electron transition metal complex oxides, which also host strong SOC, remain largely unreported. In addition to strong SOC, these complex oxides can also provide the additional tuning knob of epitaxy to control the electronic structure and the engineering of spin-to-charge conversion by crystalline symmetry. Here, we demonstrate room-temperature generation of spin-orbit torque on a ferromagnet with extremely high efficiency via the spin-Hall effect in epitaxial metastable perovskite SrIrO₃. We first predict a large intrinsic spin-Hall conductivity in orthorhombic bulk SrIrO₃ arising from the Berry curvature in the electronic band structure. By manipulating the intricate interplay between SOC and crystalline symmetry, we control the spin-Hall torque ratio by engineering the tilt of the corner-sharing oxygen octahedra in perovskite SrIrO₃ through epitaxial strain. This allows the presence of an anisotropic spin-Hall effect due to a characteristic structural anisotropy in SrIrO₃ with orthorhombic symmetry. Our experimental findings demonstrate the heteroepitaxial symmetry design approach to engineer spin-orbit effects. We therefore anticipate that these epitaxial 5d transition-metal oxide thin films can be an ideal building block for low-power spintronics.**

**spin-Hall effect | spin-orbit torque | complex oxides | strong spin-orbit coupling**

Current-induced spin-torque originating from spin-orbit effects offers an energy-efficient scheme for the electrical manipulation of magnetic devices(1–4). A large spin-torque efficiency, arising from either the spin-Hall effect(5–7) or the Rashba-Edelstein effect(8), is highly-desirable for enabling broad applications in spintronics. Great effort has been focused therein on semiconductors(9, 10), heavy metals(11–18), oxides(19, 20), and, more recently, topological insulators with a spin-momentum locked surface state(21–24). In the case of 5d-electron transition metal oxides, the delicate interplay between strong spin-orbit coupling and electron-electron correlation is believed to lead to non-trivial quantum phases(25–30), e.g., topological semimetals(31), quantum spin-Hall materials(32), and topological Mott insulators(27). The interplay between crystal structure and strong spin-orbit coupling (SOC) of conduction electrons can produce a large Berry curvature that gives rise to an intrinsic spin-Hall effect(33). Furthermore, due to the sensitive dependence of the electronic structure on crystalline symmetry, the efficiency of the spin/charge current conversion can be possibly modulated through heteroepitaxy. However, such material systems remain largely unexplored in the field of spin-orbitronics.

We first motivate our work with theoretical calculations that show a large intrinsic spin-Hall effect in the orthorhombic bulk semimetal SrIrO$_3$ (Fig. 1A). Unexpectedly large spin-Hall conductivity [SHC ~2×10$^4$ (ℏ/e)(Ω m)$^{-1}$] is obtained at the charge neutral point from the linear response theory for the bulk orthorhombic perovskite structure (see *Materials and Methods* for details). The spin-Hall effect arises due to the Berry curvature of the electronic band structure, mainly from the nearly degenerate energy bands. As shown in the momentum-resolved SHC (bottom of Fig. 1B), the high intensities of SHC appear around the *k* points where the Fermi level crosses the nearly degenerate bands. Such a characteristic band structure is unique and occurs as a combined effect of SOC and the IrO$_6$ octahedral tilting/rotation in the bulk system, which is closely related to the non-symmorphic symmetry of the bulk crystal structure(34, 35). Due to the extended nature of the SrIrO$_3$ 5d orbitals, the electronic structure of SrIrO$_3$ is sensitive to changes in lattice symmetry. We reiterate that the large spin-Hall effect occurs only when the film has the orthorhombic structure due to its origins from the underlying bulk band structure. This suggests a new approach to engineer the spin-Hall effect by precise control of oxygen octahedral tilting, which allows fine-tuning of the electronic band structure.

We illustrate our crystalline symmetry design principles for engineering a large spin-Hall effect in Fig. 1C, in which the SrIrO$_3$ crystalline symmetry is manipulated by utilizing a lattice symmetry-mismatched SrIrO$_3$/SrTiO$_3$ heterointerface and atomic layer-by-layer control of the SrIrO$_3$ layer thickness. The crystalline symmetry of perovskite oxides is directly related to the connectivity of corner-sharing octahedra that can be manipulated by epitaxial strains or interfacial couplings(36). In the case of SrIrO$_3$ films near the SrIrO$_3$/SrTiO$_3$ interface, the IrO$_6$ octahedral tilting follows the non-tilted SrTiO$_3$ substrate due to the structure imprint of the cubic symmetry of SrTiO$_3$. Such a nearly complete suppression of IrO$_6$ octahedral tilting in ultrathin SrIrO$_3$ films gives rise to a tetragonal symmetry, whereas the degree of tilting in increasingly thicker films approaches that of bulk SrIrO$_3$. Experimentally, our synchrotron x-ray diffraction of variable thickness films directly confirmed such a structural symmetry transition in SrIrO$_3$, while our measurements of the spin-Hall effect in SrIrO$_3$ showed an enhancement, coincident with this transition to orthorhombic symmetry. We also observed a large anisotropy in the spin-Hall effect enabled by the structural anisotropy of the orthorhombic state. These results demonstrate that oxide perovskites not only possess a large spin-Hall effect at room temperature, but also enables further tuning of spin-orbit effects by symmetry design.

Epitaxial SrIrO$_3$ thin films were grown on (001) SrTiO$_3$ substrates by pulsed laser deposition with *in situ* high-pressure RHEED (see *SI Appendix* for details). Ferromagnetic Permalloy Ni$_{81}$Fe$_{19}$ (Py) polycrystalline thin films were then sputtered *in situ* on SrIrO$_3$, preserving critical interface transparency for spin-current transmission and efficient spin-orbit torque (SOT) generation[37, 38]. A 1 nm $Al_2O_3$ layer was added to prevent oxidation of Py. A control sample with an *ex situ* Py/SrIrO$_3$ interface showed a much smaller efficiency for spin-torque generation. Atomic force microscopy images of the 1 nm $Al_2O_3$/3.5 nm Py/8 nm (20 unit cell) SrIrO$_3$ surface reveal an atomically-smooth surface (see *SI Appendix* for details). In Fig. 2, we show the cross-sectional filtered STEM-HAADF image of a 20 unit cell (uc) SrIrO$_3$ film on (001) SrTiO$_3$ capped with 2.5 nm Py. From the image, we determined that the high-quality SrIrO$_3$ film shares the same pseudocubic epitaxial arrangement as the SrTiO$_3$ substrate, with sharp interfaces between both SrTiO$_3$/SrIrO$_3$ and SrIrO$_3$/Py. We confirmed by x-ray diffraction that the SrIrO$_3$ film grows along $[110]_o$ (subscript o for orthorhombic notation) out-of-plane, and with $[1\bar{1}0]_o$ and $[001]_o$ in-plane along the $[100]$ and $[010]$ directions of SrTiO$_3$, respectively (see *SI Appendix* for details). For simplicity, we use pseudocubic indices $a$, $b$ and $c$ to represent orthorhombic $[1\bar{1}0]_o$, $[001]_o$, and $[110]_o$ orientations, respectively.

The spin-Hall effect in SrIrO$_3$ was probed by measuring the spin-orbit torque produced in the adjacent Py layer with spin-torque ferromagnetic resonance (ST-FMR)[12, 21, 38], as illustrated in the schematics (Fig. 3A). When an alternating charge current flows in SrIrO$_3$, the spin-Hall effect induces a spin current that flows into the Py. This spin current exerts torque on the Py and excites the magnetic moment into precession, generating an alternating change of the resistance due to the anisotropic magnetoresistance (AMR) in Py. We measure a dc voltage signal $V_{mix}$ across the device bar that arises from the mixing between the alternating current and changes in the device resistance. The ST-FMR spectrum can be obtained by sweeping external in-plane magnetic fields through the Py resonance condition (see *SI Appendix* for details), from which the resonance lineshape has been used to evaluate the efficiency of the torque. This would require a precise calibration of the microwave current in the bilayer. However, we find that the impedance of the SrTiO$_3$ substrates decreases at microwave frequencies to become comparable to the bilayer, thus leading to microwave current shunting through substrates. This may affect the result based on ST-FMR lineshape analysis (see *Materials and Methods* for details).

We evaluated the spin-torque ratio, which describes the efficiency of in-plane component of torque $\tau_\parallel$ generation relative to the charge current density in SrIrO$_3$ via spin-Hall effect ($\theta_\parallel = (\hbar/2e)j_S/j_C$, where $j_S$ is the spin current density absorbed by the Py, and $j_C$ is the applied charge current density in SrIrO$_3$), in SrIrO$_3$ by measuring the dc current-induced changes in of the ST-FMR lineshape (dc-tuned ST-FMR)[12, 38]. The injection of the dc current exerts an additional dc spin-torque on the adjacent Py, which modifies the Py resonance linewidth $W$ (and effective Gilbert damping $\alpha_{eff}$), as this torque component adds to or subtracts from the Gilbert damping torque depending on the relative orientation between the current and magnetic field[11]. This linewidth modification due to the applied dc current should be unaffected by microwave frequency current shunting through SrTiO$_3$ substrates. A quantitative analysis of the spin-torque ratio $\theta_\parallel$ for 3.5 nm Py/8 nm (20 uc) SrIrO$_3$ bilayers is shown in Fig. 3B, where $W$ and $\alpha_{eff}$ scale linearly with the applied dc current. The current is applied parallel to the $a$-axis and the in-plane magnetic field is swept at an angle $\varphi = -45°$, with respect to the current axis. The magnitude of the in-plane torque is proportional to the change of the effective Gilbert damping $\alpha_{eff}$ over the current density $j_c$ in SrIrO$_3$. By averaging measurements at different frequencies, we find that the spin-torque ratio, $\theta_\parallel = 0.51\pm0.07$ for 3.5 nm Py/8 nm (20 uc) SrIrO$_3$, is higher than any value reported for heavy metal systems[7, 39], and 4 times as larger as that of a 4 nm Py/ 4 nm Pt control sample ($\theta_\parallel = 0.12\pm0.03$). The sign for $\theta_\parallel$ is consistent with our first-principle calculations for bulk SrIrO$_3$ and experimental results on the heavy metal Pt. We examine the symmetry of the in-plane current-induced torque by performing Py magnetization angular dependence dc-tuned ST-FMR. Fig. 3C shows the current-induced change in the effective damping $\Delta\alpha_{eff}/j_c$ (slope of the linear fit in Fig. 3B) as a function of in-plane magnetic field angle $\varphi$, which fits to $\sin\varphi$. This is consistent with the symmetry of the conventional current-induced torque that has been observed previous in polycrystalline heavy metal/ferromagnet bilayer systems(15).

As a comparison, a significantly smaller spin torque ratio $\theta_\parallel$ was observed in the 3.5 nm Py/3.2 nm (8 uc) SrIrO$_3$ bilayer sample as shown in Fig. 3D where the current-induced change in the effective damping is smaller. We attribute the large difference of $\theta_\parallel$ in 8 uc and 20 uc SrIrO$_3$ samples to their structural transition with thickness (from tetragonal to orthorhombic) due to the structural imprint of non-tilted SrTiO$_3$ substrates as illustrated in Fig.1C. This agrees with our theory prediction of large spin-Hall effect in SrIrO$_3$ with an orthorhombic structure (the dependence of the spin torque ratio on SrIrO$_3$ film thickness will be discussed next).

Interestingly, we also observed a significant dependence of the spin-torque ratio on crystallographic orientation in the thicker SrIrO$_3$ sample (3.5 nm Py/8 nm SrIrO$_3$) as shown in Fig. 3E (blue circle), where the $\theta_\parallel$ was extracted from devices with the charge current applied along various in-plane crystal orientations (from $a$- to $b$-axis) while keeping the magnetic field angle $\varphi=-45°$. Owing to the anisotropic structural characteristics of the orthorhombic symmetry, the IrO$_6$ octahedral rotation is out-of-phase along the $a$-axis (top-left of Fig. 3E), whereas it is in-phase along the $b$-axis (top-right of Fig. 3E), which gives rise to a different oxygen octahedral configuration when the charge current is applied along the $a$- or $b$-axis, and, therefore, the anisotropic $\theta_\parallel$. This contrasts the nearly isotropic $\theta_\parallel$ observed in the thinner SrIrO$_3$ sample (3.5 nm Py/8 nm SrIrO$_3$, red square), which was found to exhibit almost total suppression of the octahedral tilt along both the $a$- and $b$-axes (bottom of Fig. 3E) compared to the thicker film (Supplmentary Information). Such dependence on crystallographic orientation of the spin torque ratio in tetragonal and orthorhombic SrIrO$_3$ demonstrates the strong correlation between the spin-Hall effect and the crystalline symmetry.

The dependence of the spin-torque ratio on the thickness of the SrIrO$_3$ layer and on the direction of current relative to the crystal axes was also confirmed by another independent characterization technique, in which the spin-torque-induced magnetization rotation in Py was measured by the polar magneto-optic-Kerr-effect (MOKE)(40). In a 5.5 nm Py/10 nm SrIrO$_3$ sample with the current applied along $a$-axis, the spin-torque ratio was measured to be 0.15±0.01, which is 2-3 times as large as a Py/Pt control sample measured by the MOKE (0.065±0.01, see *SI Appendix* for details). This is qualitatively consistent with our dc-tuned ST-FMR results, although the absolute value of the spin-torque ratio is smaller. Nevertheless, by using the MOKE, we also observe the large lattice symmetry dependence ($\theta_\parallel$=0.15±0.01 in the 25 uc SrIrO$_3$ sample, while 0.013±0.01 in the 8 uc SrIrO$_3$ sample) and anisotropic spin-torque ratio ($\theta_\parallel$=0.15±0.01 with $j_c \| a$-axis, while 0.05±0.005 with $j_c \| b$-axis in the 25 uc SrIrO$_3$ sample), which are qualitatively consistent with the trends obtained in our dc-tuned ST-FMR measurements.

To analyze the effect of lattice symmetry on spin-Hall effect in detail, we studied the dependence of the spin-torque ratio on varying thicknesses of SrIrO$_3$ films. From synchrotron x-ray measurements, we robustly established the previously-discussed suppression of IrO$_6$ octahedral tilt for ultrathin SrIrO$_3$ films, which modifies the symmetry of SrIrO$_3$ from orthorhombic to tetragonal (see *SI Appendix* for details)(41–43). As shown in Fig. 4A, this transition is characterized by the orthorhombicity factor defined as $a_o/b_o$ (orthorhombic indices), which illustrates a tetragonal-to-orthorhombic SrIrO$_3$ structural transition around 4.8 nm (12 uc). We then measured the spin-torque ratio on the same series of SrIrO$_3$/Py bilayer samples (with fixed Py thickness). Accordingly, as shown in Fig. 4B, $\theta_\parallel$ increases sharply and saturates at the thickness of 6.4 nm (16 uc) from a nearly constant value when the thickness is below 10 uc. The spin-torque ratio is also expected to increase and saturate when the thickness of the spin-Hall source material exceeds the spin diffusion length based on the standard spin diffusion theory(44). However, the abrupt change of $\theta_\parallel$ occurs at the SrIrO$_3$ structural transition thickness, which cannot be explained simply by the diffusive spin transport, as we determined a short spin diffusion length of ~1.4 nm (3.5 uc) in SrIrO$_3$ from the thickness dependence of the over-layer Py Gilbert damping enhancement (Fig. 4C, see *Materials and Methods* for details). This strong dependence of $\theta_\parallel$ on lattice symmetry rather than on spin diffusion illustrates a direct connection between the degree of IrO$_6$ octahedral tilt and the spin-torque efficiency.

In summary, we have proposed and developed a new material for spin-orbit torque applications in a transition metal perovskite with spin-orbit coupled *5d* electrons, in which the interplay between SOC and the crystal structure produces a large spin-torque ratio. Furthermore, the extended nature of *5d* orbitals allows a sensitive response of the electronic band structure to an externally manipulated lattice structure, as manifested in the strong dependency of the spin-Hall effect on the degree of IrO$_6$ octahedral tilting in epitaxially-strained SrIrO$_3$. Such intricate coupling between the electronic and lattice degrees of freedom can thus open up a new avenue to engineer spin-orbit effects by tailoring the lattice symmetry through heteroepitaxy. This material acts as an ideal building block for oxide spintronics, since a broad range of ferromagnetic perovskites can be integrated in an epitaxial heterostructure with atomically sharp interfaces for efficient spin-current transmission. Therefore, we anticipate that the use of *5d* transition metal perovskites could lead to substantial advances in spintronics.

## Materials and Methods

**Sample growth, fabrication and characterization.** SrIrO$_3$ films were epitaxially synthesized on (001) SrTiO$_3$ substrates using pulsed laser deposition (PLD). During the growth, layer-by-layer deposition was observed by *in situ* reflection high energy electron diffraction (RHEED). Before the growth, the SrTiO$_3$ (001) substrates were chemically etched and annealed to ensure TiO$_2$ surface termination. The substrates were first immersed in buffered hydrofluoric acid for 60 seconds before being annealed at 900 °C for 6 hours in an O$_2$-rich environment. After annealing, the substrates were etched again in buffered hydrofluoric acid to remove any leftover SrO on the surface. The PLD growth was conducted at a substrate temperature of 600 °C and an oxygen partial pressure of 75 mTorr. The laser fluence at the SrIrO$_3$ target surface was ~1 J/cm$^2$ and the pulse repetition was 10 Hz. The working distance between target and substrate was ~58 mm. After the SrIrO$_3$ growth, the sample was cooled down in an oxygen rich atmosphere. The chamber was re-evacuated at room temperature and Py was sputter deposited at an Ar pressure of 3 mTorr with a background pressure <2×10$^{-8}$ Torr, followed by a 1 nm Al passivation layer. The Py film is shown to be polycrystalline, which we confirmed by the observation of RHEED diffraction rings after deposition. The atomically flat Py film on top of SrIrO$_3$ was verified using atomic force microscopy (see *SI Appendix* for details). We confirmed the thickness, epitaxial arrangement, and coherence of the SrIrO$_3$ films using x-ray reflectivity, x-ray diffraction, and reciprocal space mappings (see *SI Appendix* for details). The coherent SrIrO$_3$ films assembled on the cubic SrTiO$_3$ substrate with the orthorhombic SrIrO$_3$ [1-10]$_o$, [001]$_o$, and [110]$_o$ directions along the cubic SrTiO$_3$ [100], [010], and [001] directions, respectively, and the misfit epitaxial strain from the SrTiO$_3$ induced a monoclinic distortion of the SrIrO$_3$ unit cell, which is consistent with similar orthorhombic perovskites grown on cubic substrates (see *SI Appendix* for details)(45). The thickness of Py films was measured by using x-ray reflectivity.

We patterned the Py/SrIrO$_3$ sample by using photolithography followed by ion beam milling. Then 200 nm Pt/5 nm Ti electrodes were sputter deposited and defined by a lift-off procedure. Devices for ST-FMR were patterned into microstrips (20-50 μm wide and 40-100 μm long) with ground-signal-ground electrodes. Devices for electrical transport measurements were patterned into 100 μm wide and 500 μm long Hall bars.

**STEM measurements.** TEM specimens were prepared by a focused ion multi-beam system (JIB-4610F, JEOL, Japan). To protect the Py/SrIrO$_3$ films, an amorphous carbon layer was deposited on the top surface before the ion beam milling. A Ga$^+$ ion beam with an acceleration voltage of 30 kV was used to fabricate the thin TEM lamella. To minimize the surface damage induced by the Ga$^+$ ion beam milling, the sample was further milled by an Ar$^+$ ion beam (PIPS II, Gatan, USA) with an acceleration voltage of 100 meV for 4 minutes. HAADF-STEM images were taken by using a scanning transmission electron microscope (JEM-2100F, JEOL, Japan) at 200 kV with a spherical aberration corrector (CEOS GmbH, Germany). The optimum size of the electron probe was ~0.9 Å. The collection semi-angles of the HAADF detector were adjusted from 70 to 200 mrad in order to collect large-angle elastic scattering electrons for clear $Z$-sensitive images. The obtained raw images were processed with a band-pass Wiener filter with a local window to reduce a background noise (HREM research Inc., Japan).

**ST-FMR measurements.** During ST-FMR measurements, a microwave current at a fixed frequency (4.5 to 7 GHz) is applied through the ac port of a bias-T to a RF ground-signal-ground probe tip. The microwave power output (8 to 14 dBm) is also fixed. The in-plane magnetic fields are generated by a rotary electromagnet which allows for magnetic field angle dependence of ST-FMR measurements. Magnetic fields are swept from 0-0.12 T for driving the Py through its resonance condition. The resonance line shape can be fitted to a sum of symmetric and antisymmetric Lorentzian components, where the anti-damping (in-plane, $\tau_\parallel$) and field-like torque (out-of-plane, $\tau_\perp$) components are proportional to the amplitudes of the symmetric and antisymmetric line shape, respectively. By performing the lineshape analysis, the spin-torque ratio can be determined. However, we find through RF transmission measurements that although bare SrTiO$_3$ is nonconductive at low frequencies, after processing, it has a finite impedance at microwave frequencies, and as a consequence RF current is shunted through the SrTiO$_3$ substrate during ST-FMR measurements. This shunting may cause potential additional contributions to the Oersted field and modify the size of the antisymmetric component of the lineshape. Additionally, the unknown current shunted through the SrTiO$_3$ poses a challenge for determining the RF current flowing through our SrIrO$_3$ layer based on RF transmission calibration. Hence, it is difficult to determine the size of the spin-Hall effect in SrIrO$_3$ from RF lineshape analysis alone. For this reason, we rely on dc-tuned ST-FMR measurements, for which the linewidth modification due to the applied DC current should be unaffected by microwave frequency current shunting, and polar magneto-optic Kerr effect (MOKE) measurements, for which RF current is unnecessary. Both of these measurement methods should be unaffected by the high-frequency conductivity of the SrTiO$_3$ substrate.

In dc-tuned ST-FMR, we modulate the rf current amplitude and measure the mixing voltage signal by using a lock-in amplifier through the dc port of the bias-T. The modulation frequency is 437 Hz. The resonance line shape is fitted to a sum of symmetric and antisymmetric Lorentzian components, from which the resonance linewidth is extracted. We quantify the $\theta_\parallel$ by linear fitting the current dependent resonance linewidth or $\alpha_{eff}$ as $|\theta_\parallel| = \frac{2|e|}{\hbar} \frac{(H_{FMR} + M_{eff}/2)\mu_0 M_s t_{FM}}{|sin\varphi|} |\frac{\Delta\alpha_{eff}}{\Delta j_c}|$, where $\alpha_{eff}$ is the effective magnetic damping of Py that is related to $W$ as $\alpha_{eff} = \frac{\gamma}{2\pi f} W$; and $\gamma, \mu_0, M_{eff}, M_S$, and $t_{FM}$ are the gyromagnetic ratio, the permeability in vacuum, the effective magnetization, the saturation magnetization and the thickness of Py, respectively. The charge current density $j_c$ is carefully calibrated by measuring the 4-point-resistance for each layer with a parallel resistor model (see *SI Appendix* for details).

**MOKE measurements.** Polar MOKE measurements of $\theta_\parallel$ were performed using a 630 nm diode laser, a Glan-Taylor polarizer with an extinction coefficient of $10^5$:1, and a 10× infinity corrected microscope objective to focus 2 mW of optical power down to a 10 μm spot. An excitation current through the device of 7-10 mA at 5667 Hz was used to lock in to the response of the magnetization due to the anti-damping spin torque. The laser spot was scanned across the middle of the device at external fields of 0.08 T parallel and anti-parallel to the current flow direction in the device. The linear polarization incident on the sample was 45° rotated from the external field direction to suppress quadratic MOKE effects. $\theta_\parallel$ is quantified as by using the ratio of the integrated absolute value of the signal across the device that is even under the external magnetic field reversal ($A_{sum}$) and odd under magnetic field reversal ($A_{diff}$) according to $\theta_{\parallel} = \frac{A_{diff}}{A_{sum}} \frac{e}{\hbar} \frac{\mu_0 M_s t_{FM} t_{tot} ln(4)}{X\pi}$ (see *SI Appendix* for details), where $t_{tot}$ is the total thickness of the $SrIrO_3$/Py bilayer and $X$ is fraction of current flowing through the $SrIrO_3$ as determined via the parallel resistor model described in the *SI Appendix*.

**Synchrotron X-ray Thin Film diffraction.** Synchrotron X-ray diffraction measurements were carried out to precisely characterize the structural and lattice symmetry evolution as a function of thickness of $SrIrO_3$ thin films epitaxially grown on a (001) $SrTiO_3$ substrate. The thin film diffraction measurements were performed on a five-circle diffractometer with χ-circle geometry, using an X-ray energy of 20 keV (wavelength λ = 0.6197 Å) at sector 12-ID-D of the Advanced Photon Source, Argonne National Laboratory. The X-ray beam at the beamline 12-ID-D has a total flux of $4.0 \times 10^{12}$ photons/s and was vertically focused by beryllium compound refractive lenses down to a beam profile of ∼ 50 μm. The L-scans along respective truncation rods {10L} were obtained by integrating the diffuse background contributions using the two-dimensional images acquired with a pixel 2D array area detector (Dectris PILATUS-1mm Si 100K). The separation of respective {103} film peak positions in reciprocal space can be used to extract the out-of-plane tilt angle of the $SrIrO_3$ film with respect to the cubic $SrTiO_3$ lattice, so that we can obtain the degree of orthorhombic distortion (a/b> 1) for each $SrIrO_3$ thin film as a function of thickness (see *SI Appendix* for details).

**Estimation of the spin diffusion length in Py/$SrIrO_3$.** To estimate the spin diffusion length in our $SrIrO_3$ thin film, we characterize the Gilbert damping parameter $\alpha$ of Py in Py/$SrIrO_3$ bilayers with various $SrIrO_3$ thicknesses by using both ST-FMR (on patterned samples) and broadband FMR measurements (on 5 mm by 5 mm samples). In broadband FMR measurements, the microwave magnetic field is produced by a coplanar waveguide, and the resonance spectrum is obtained by sweeping external magnetic fields through the Py resonance condition with a fixed microwave frequency. We measure the field derivative of the FMR absorption intensity with field modulation at low frequency (<1 kHz). In both techniques, we perform frequency dependent measurements, from which the resonance linewidth of Py is obtained at each frequency. The Gilbert damping parameter $\alpha$ is then calculated as $W = W_0 + \frac{2\pi\alpha}{|\gamma|}f$, where $W_0$ is the inhomogeneous linewidth broadening. We observed the enhancement of $\alpha$ with increasing $SrIrO_3$ thickness effect as shown in Fig. 4C. The data can be fitted to a diffusive spin transport model as(46), $\alpha = \alpha_0 + \frac{g_{op}\mu_B \hbar}{2e^2 M_s t_{SIO}}[\frac{1}{G_{\uparrow\downarrow}} + 2\rho(t)\lambda_s \coth(\frac{t}{\lambda_s})]^{-1}$, where $g_{op}$ is the Lande $g$ factor, $\alpha_0$ is the Gilbert damping with zero $SrIrO_3$ thickness, $G_{\uparrow\downarrow}$ is the interfacial spin mixing conductance per unit area, $t$ is the thickness of $SrIrO_3$, $\rho(t)$ is the thickness dependent resistivity of $SrIrO_3$ and $\lambda_s$ is the spin diffusion length in $SrIrO_3$. In this fitting, we take into account the thickness dependence of the $SrIrO_3$ resistivity which increases in ultrathin films due to scattering mechanisms (see *SI Appendix* for details), which gives a spin mixing conductance $G_{\uparrow\downarrow}$ of $1.8 \times 10^{14}$ $\Omega^{-1}$ $m^{-2}$ and a spin diffusion length of 1.4 nm. This is consistent with that reported in the $LaSrMnO_3$/$SrRuO_3$ system(47).

**Theoretical calculations.** For the spin-Hall conductivity calculations, we employed a $j_{eff}$=1/2 tight-binding model constructed for the orthorhombic perovskite bulk $SrIrO_3$(34, 35). The model incorporates various spin-dependent hopping channels for Ir electrons generated by oxygen octahedron tilting in the bulk structure. The model Hamiltonian $H$ consists of four doubly degenerate electron bands on account of the four Ir sites in each unit cell.

$$H = \sum_{\boldsymbol{k}} \psi_{\boldsymbol{k}}^{\dagger} H_{\boldsymbol{k}} \psi_{\boldsymbol{k}}$$

Here, $\psi = (\psi_{1\uparrow}, \psi_{2\uparrow}, \psi_{3\uparrow}, \psi_{4\uparrow}, \psi_{1\downarrow}, \psi_{2\downarrow}, \psi_{3\downarrow}, \psi_{4\downarrow})^T$ are electron operators with the subscripts meaning the sub-lattice (1,2,3,4) and $j_{eff}$=1/2 pseudo-spin ($\uparrow, \downarrow$). The explicit form of $H_k$ and the values of the hopping parameters can be found in Refs. 24 and 25. The electron band structure of the model is displayed in Fig. 1B. Then, the SHC tensor $\sigma_{\mu\nu}^{\rho}$ is calculated by the Kubo formula(7, 48):

$$\sigma_{\mu\nu}^{\rho} = \sum_{\boldsymbol{k}} \Omega_{\mu\nu}^{\rho}(\boldsymbol{k})$$

where

$$\Omega_{\mu\nu}^{\rho}(\boldsymbol{k}) = \frac{2e\hbar}{V} \sum_{\epsilon_{\mathrm{nk}} < \epsilon_F < \epsilon_{\mathrm{mk}}} \mathrm{Im}\left[\frac{\langle \mathrm{mk}|\mathcal{J}_{\mu}^{\rho}|\mathrm{nk}\rangle\langle \mathrm{nk}|J_{\nu}|\mathrm{mk}\rangle}{(\epsilon_{\mathrm{mk}} - \epsilon_{\mathrm{nk}})^2}\right].$$

Here, $J_{\nu}(= \sum_{\boldsymbol{k}} \psi_{\boldsymbol{k}}^{\dagger} \frac{\partial H_{\boldsymbol{k}}}{\partial k_{\nu}} \psi_{\boldsymbol{k}})$ is charge current, and $\mathcal{J}_{\mu}^{\rho}(= \frac{1}{4}\{\sigma^{\rho}, J_{\mu}\})$ is spin current with the $j_{eff}$=1/2 spin represented by the Pauli matrix $\sigma^{\rho}$. In the above expression, $V$ is the volume of the system, $\epsilon_F$ is the Fermi level, and $|\mathrm{mk}\rangle$ represents a Bloch state of $H$ with energy $\epsilon_{\mathrm{mk}}$. The momentum-resolved SHC represented by $\Omega_{\mu\nu}^{\rho}(\boldsymbol{k})$ enables us to trace the electron states responsible for the large spin-Hall effect. More details of the theoretical calculation of SHC in SrIrO$_3$ can be found in a recent theoretical paper(49) by 2 authors from the current work .

**ACKNOWLEDGMENTS.** We acknowledge E. Y. Tsymbal, P. J. Ryan, D. D. Fong, and J. Irwin for discussions, S. Emori for comments on the manuscript, and Y. J. Ma, D. T. Harris, L. Guo, and J. Schad for technical assistance with experiments. T. Nan acknowledges L. C. Sun for assistance in graphic design. This work was supported by the National Science Foundation under DMREF Grant No. DMR-1629270, AFOSR FA9550-15-1-0334 and AOARD FA2386-15-1-4046. Work at Northeastern University is funded by the NSF TANMS ERC Award 1160504. This research used resources of the Advanced Photon Source, a U.S. Department of Energy (DOE) Office of Science User Facility operated for the DOE Office of Science by Argonne National Laboratory under Contract No. DE-AC02-06CH11357. Work at Cornell was supported by the National Science Foundation (DMR-1406333) and Western Digital, and made use of the Cornell Center for Materials Research Shared Facilities which are supported by the NSF MRSEC program (DMR-1120296). K. Hwang and Y. B. Kim are supported by the NSERC of Canada, Canadian Institute for Advanced Research, and Center for Quantum Materials at the University of Toronto.

rotations and lattice modulations. *Phys Rev B* 83:064101.

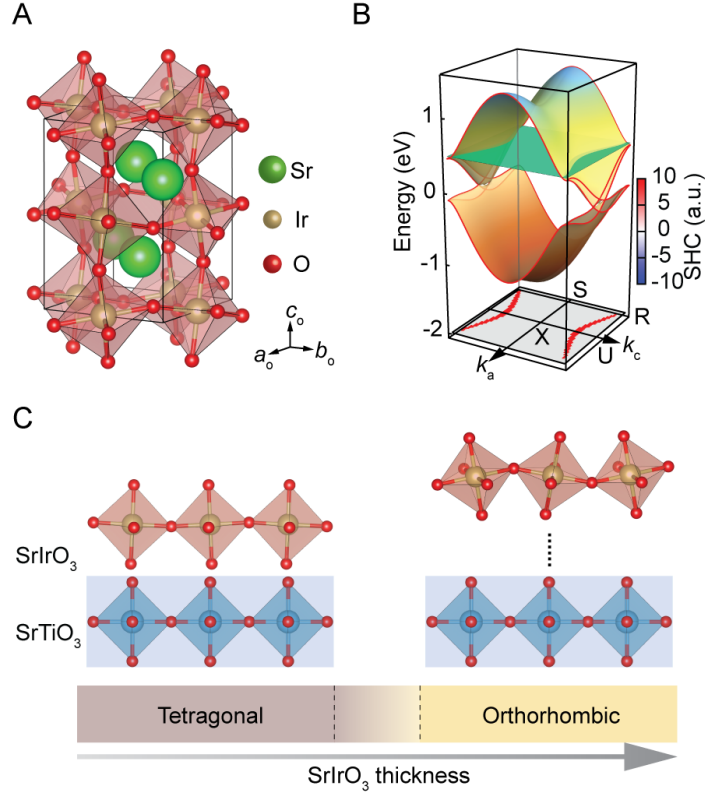

**Fig. 1.** The mechanism of the spin-Hall effect in SrIrO$_3$. (*A*) Orthorhombic perovskite crystal structure of bulk SrIrO$_3$, where $a_o$, $b_o$, and $c_o$ correspond to the [100]$_o$, [010]$_o$, and [001]$_o$ directions (subscript o for orthorhombic notation), respectively. (*B*) Electron energy band structure of the bulk system on the XURS plane of the Brillouin zone and the momentum-resolved SHC $\Omega(k)$ in the Brillouin zone at the charge neutrality point. The red arcs at the bottom represent the momentum-resolved SHC, which is the net Berry curvature summed over occupied electron levels below a given Fermi level (green). (*C*) Schematic illustrations of the lattice symmetry of SrIrO$_3$ when grown on cubic non-tilted SrTiO$_3$, where a partial suppression of the IrO$_6$ octahedral tilt exists up to a certain thickness before a tilted octahedral pattern is established in thicker films.

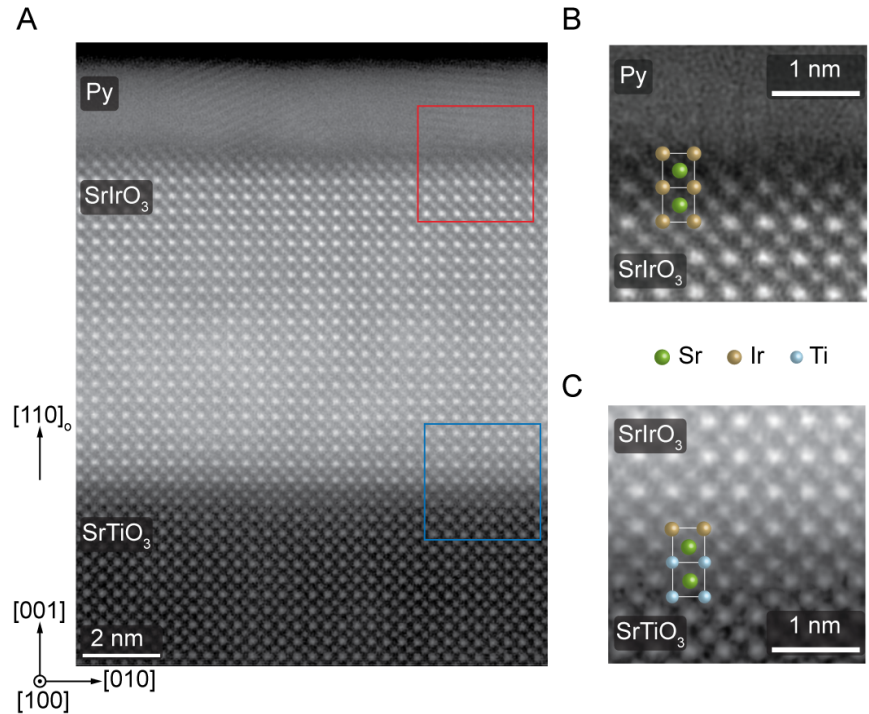

**Fig. 2.** The structural characterization of the Py/SrIrO$_3$/SrTiO$_3$ system. (*A*) Scanning transmission electron microscope image of Py/SrIrO$_3$ heterostructure on (001) SrTiO$_3$ substrate. The image contrast is approximately proportional to the atomic number $Z$, where brighter colors represent heavier elements. (*B,C*) Expanded image of the top Py/SrIrO$_3$ interface *B*, and the bottom SrIrO$_3$/SrTiO$_3$ interface *C*, showing the high quality SrIrO$_3$ film with atomically-sharp interfaces. The stacking of the atomic constituents is highlighted in the blown-up images by the superimposed filled dots of different colors (Sr in green, Ir in yellow and Ti in blue).

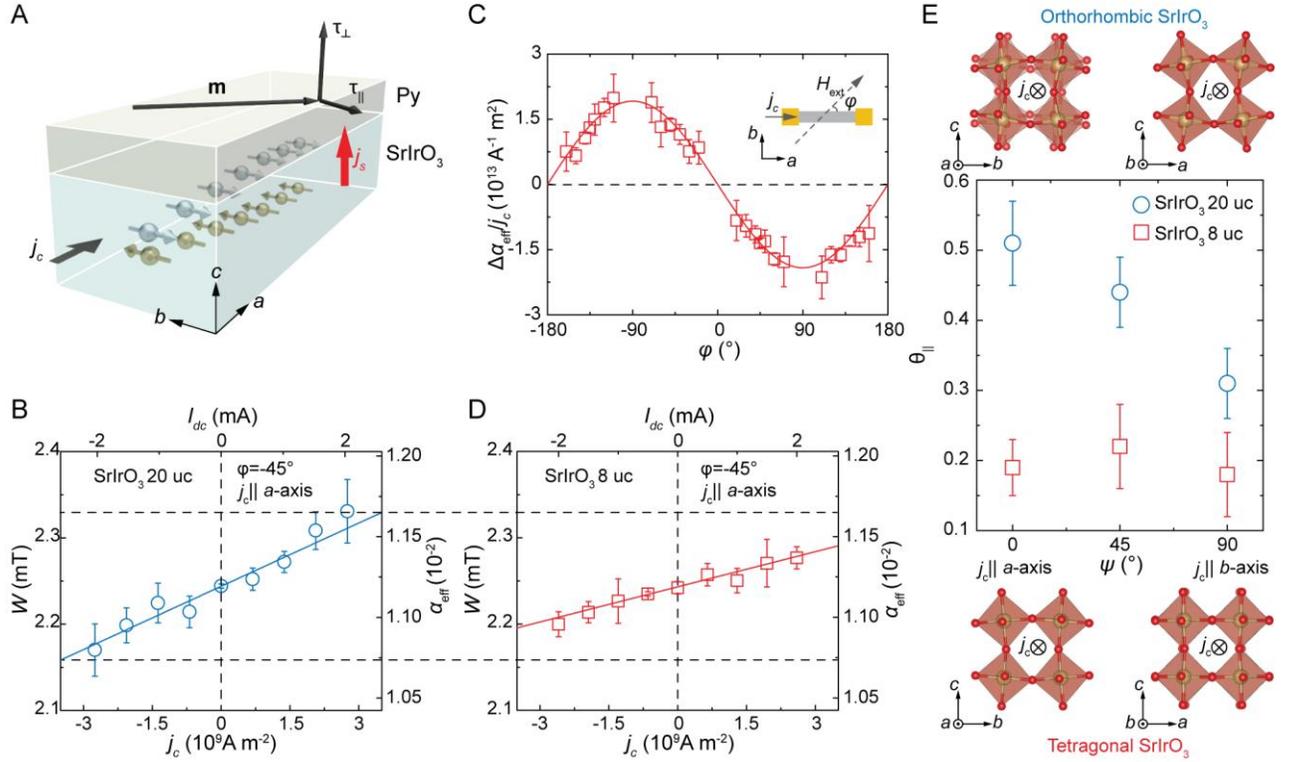

**Fig. 3.** ST-FMR measurements. (*A*) Schematic of the Py/SrIrO₃ bilayer on SrTiO₃(001) and the current-induced torque geometries. Incoming charge current ($j_c$) generates a spin current ($j_s$) along the out-of-plane [110]ₒ direction or *c*-axis. The pseudocubic indices *a*, *b* and *c* correspond to orthorhombic [1$\bar{1}$0]ₒ, [001]ₒ and [110]ₒ directions, respectively. (*B*) Resonance linewidth *W* and effective magnetic damping $\alpha_{eff}$ as functions of dc charge current $I_{dc}$ and current density $j_c$ in SrIrO₃ for a 3.5 nm Py/8 nm (20 uc) SrIrO₃ microstripe (20 μm × 40 μm). The current is applied along the [1$\bar{1}$0]ₒ direction (*a*-axis) and the external magnetic field is oriented at an angle $\varphi$=-45° with respect to the current axis. The applied microwave frequency and power are 5.5 GHz and 15 dBm, respectively. The solid line represents a linear fitting. (*C*) Current modulation of Py effective damping as a function of external magnetic field angle for the 3.5 nm Py/8 nm (20 uc) SrIrO₃ sample. The solid line shows the fit to sin($\varphi$). Resonance linewidth *W* and effective magnetic damping $\alpha_{eff}$ as functions of dc current and current density. (*D*) Change of *W* and $\alpha_{eff}$ due to $j_c$ in a 3.5 nm Py/3.2 nm (8 uc) SrIrO₃ sample. (*E*) In-plane crystallographic orientation dependence of the spin-torque ratio $\theta_\parallel$ for Py/SrIrO₃ bilayers (with Py thickness fixed at 3.5 nm) with two different SrIrO₃ film thicknesses. $\psi$ is the angle between the [1$\bar{1}$0]ₒ direction (*a*-axis) and the applied current axis. The top and bottom schematics illustrate the planar crystalline geometry of SrIrO₃ viewed along *a*- and *b*-axis for orthorhombic and tetragonal symmetries, respectively.

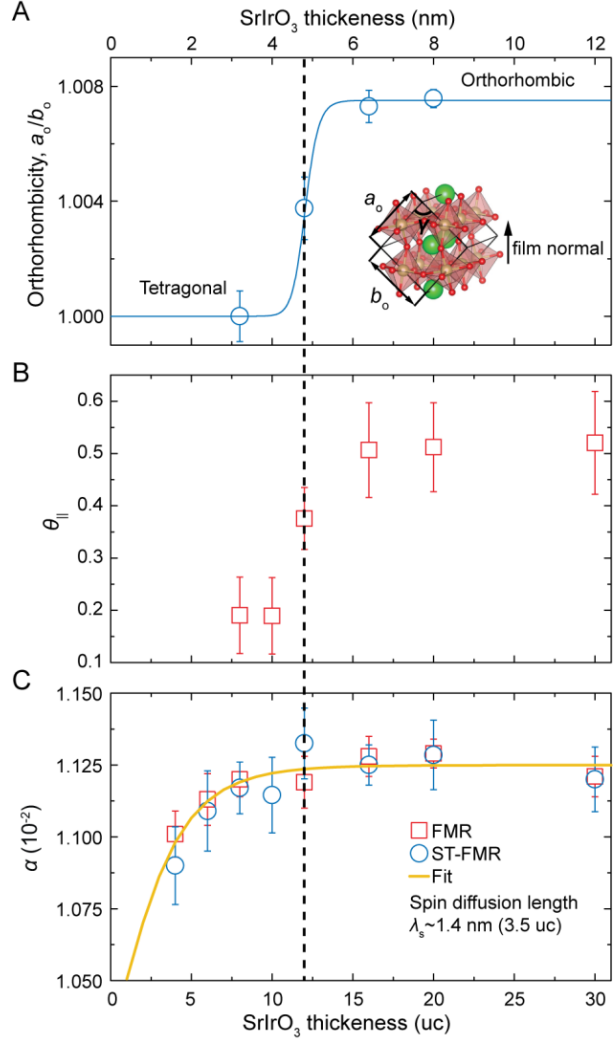

**Fig. 4.** Control of spin-torque ratio with lattice symmetry stabilization. (*A*) Thickness dependence of orthorhombicity factor, defined as $a_o/b_o$, of SrIrO$_3$ thin films. The solid line is a guide to illustrate the SrIrO$_3$ crystalline symmetry transition. Inset shows the schematic of the SrIrO$_3$ orthorhombic unit cell with the tilted IrO$_6$ octahedra, where the arrow indicates the thin film growth direction (*c*-axis or [110]$_o$). $\gamma$ is the angle between $a_o$- and $b_o$-axis. (*B*) SrIrO$_3$ thickness dependence of spin orbit ratio $\theta_\parallel$ for Py/SrIrO$_3$ bilayers (with Py thickness fixed at 3.5 nm). (*C*) SrIrO$_3$ thickness dependence of Gilbert damping parameter $\alpha$ determined by the broadband FMR (red) and ST-FMR (blue) measurements for Py/SrIrO$_3$ bilayers (with Py thickness fixed at 3.5 nm). The solid line represents the fit to the diffusive spin-pumping model, which gives a spin-diffusion length $\lambda_s$ ~1.4 nm for SrIrO$_3$.

SI Appendix for

# Anisotropic spin-orbit torque generation in epitaxial SrIrO₃ by symmetry design

T. Nan[a,1], T. J. Anderson[a,1], J. Gibbons[b], K. Hwang[c], N. Campbell[d], H. Zhou[e], Y. Q. Dong[e], G. Y. Kim[f], N. Reynolds[b], X. J. Wang[g], N. X. Sun[g], S. Y. Choi[f], M. S. Rzchowski[d], Yong Baek Kim[c,h,i], D. C. Ralph[b,j] and C. B. Eom[a,2]

[a]Department of Materials Science and Engineering, University of Wisconsin-Madison, Madison, Wisconsin 53706, USA; [b]Laboratory of Atomic and Solid State Physics, Cornell University, Ithaca, New York 14853, USA; [c]Department of Physics and Centre for Quantum Materials, University of Toronto, Toronto, Ontario M5S 1A7, Canada; [d]Department of Physics, University of Wisconsin-Madison, Madison, Wisconsin 53706, USA; [e]Advanced Photon Source, Argonne National Laboratory, Argonne, Illinois 60439, USA; [f]Department of Materials Science and Engineering, POSTECH, Pohang 37673, Korea; [g]Department of Electrical and Computer Engineering, Northeastern University, Boston, Massachusetts 02115, USA; [h]Canadian Institute for Advanced Research/Quantum Materials Program, Toronto, Ontario M5G 1Z8, Canada; [i]School of Physics, Korea Institute for Advanced Study, Seoul 130-722, Korea; and [j]Kavli Institute at Cornell for Nanoscale Science, Ithaca, New York 14853, USA

## I. Structure and surface characterization of SrIrO₃/SrTiO₃(001) heterostructure

In Fig. S1A, the RHEED intensity spectrum of a 20 uc (8 nm) SrIrO₃ film on SrTiO₃ is plotted. The intensity oscillations indicate layer-by-layer growth of the film. The RHEED pattern in the right inset indicates near preservation of the substrate RHEED pattern in the left inset. After deposition of Py onto the SrIrO₃ film, RHEED patterns of the Py surface show faint rings, indicating a textured polycrystalline Py structure. After the *in situ* Al₂O₃/Py deposition, atomic force microscopy images were taken. As can be seen in Fig. S1B, the final surface of the Al₂O₃/Py/SrIrO₃//SrTiO₃ (001) surface retains the step-terrace features of the chemically and thermally treated TiO₂-terminated SrTiO₃ substrate in Fig. S1B.

In Fig. S2, the lab-source x-ray diffraction data of a 1nm Al₂O₃/3.5 nm Py/12 nm SrIrO₃/SrTiO₃ (001) heterostructure is presented. For the purposes of the x-ray discussion in this supplement, we define pseudocubic (pc) lattice parameters for SrIrO₃, where $[100]_{pc}$, $[010]_{pc}$, and $[001]_{pc}$ lie along the SrIrO₃ orthorhombic (o) $[1\bar{1}0]_o$, $[001]_o$, and $[110]_o$ directions, respectively. The $2\theta$-$\omega$ out-of-plan scan aligned to the (002) SrTiO₃ peak shows an epitaxial SrIrO₃ film without the presence of additional peaks that would indicate that different phases of SrIrO₃ exist. The SrIrO₃ films show distinct Kiessig fringes around the main $(001)_{pc}$ and $(002)_{pc}$ pseudocubic reflections, which indicates a smooth film surface and interfacial structure. The azimuthal $\Phi$-scan around the $(101)_{pc}$ pseudocubic reflection shows that the SrIrO₃ film shares the same pseudocubic arrangement with the underlying SrTiO₃ substrate. From the reciprocal space mapping around the (103) SrTiO₃ peak shown in Fig. S2C, we show that our SrIrO₃ films are fully coherent with the underlying SrTiO₃ substrate.

---

[1]T.N. and T.J.A. contributed equally to this work.
[2]To whom correspondence may be addressed. Email:ceom@wisc.edu

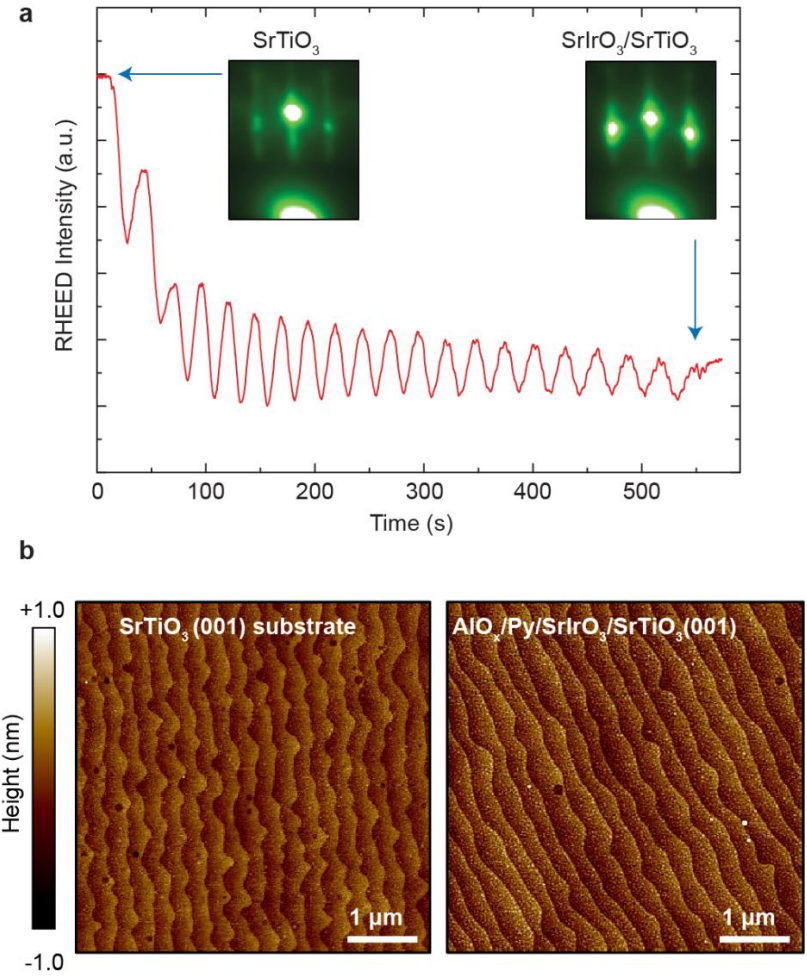

**Fig. S1.** *(A)* RHEED intensity data of SrIrO₃ growth showing clear layer-by-layer growth of a 20 uc (8nm) SrIrO₃ thin film on SrTiO₃ (001) substrate. The RHEED pattern at the end of the growth (right inset) indicates a high quality SrIrO₃ thin film with minimal surface roughening compared to that of the SrTiO₃ substrate (left inset). *(B)* Atomic force microscopy images of a treated SrTiO₃ substrate and subsequently deposited heterostructure of 1 nm AlOₓ/3 nm Py/8 nm SrIrO₃ showing near preservation of the atomically smooth SrTiO₃ substrate surface.

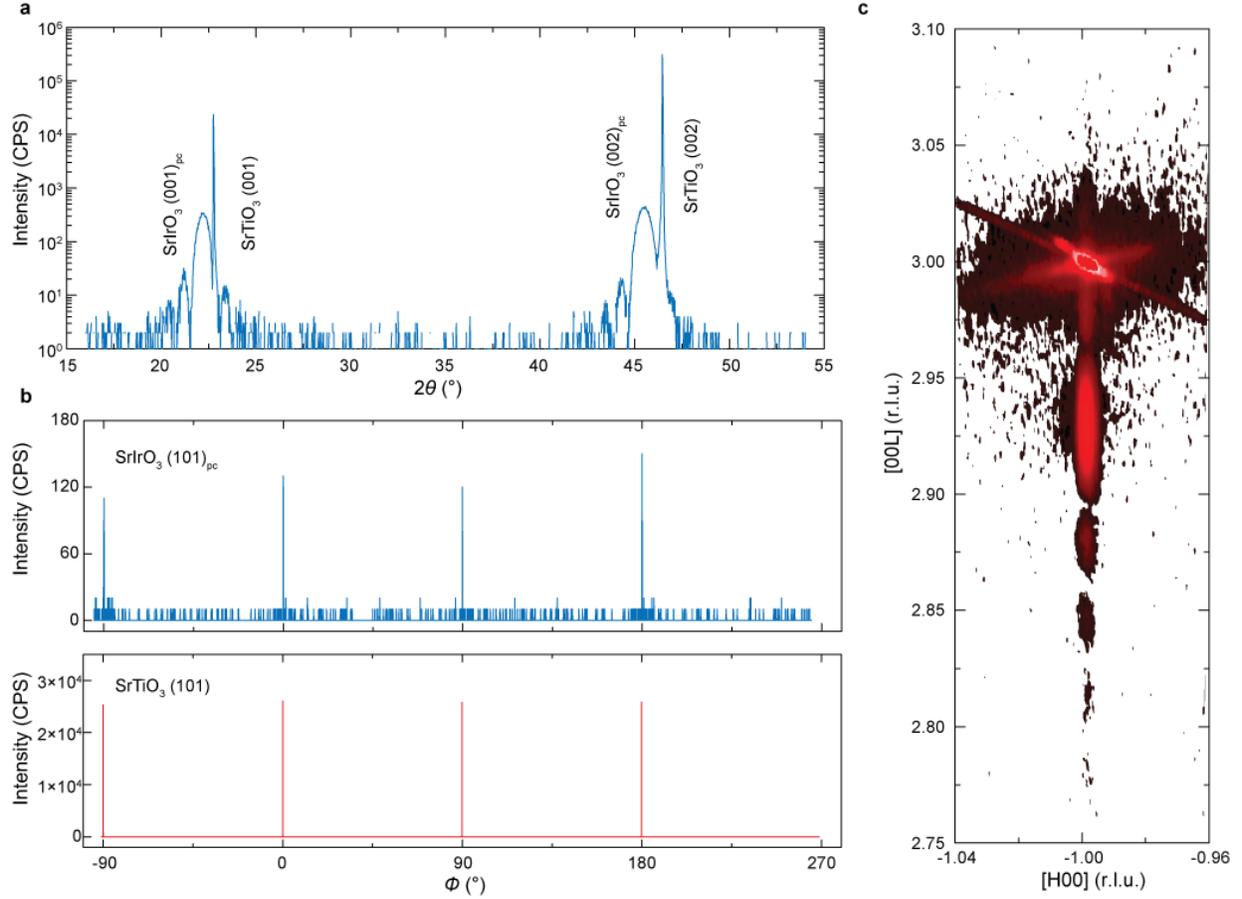

**Fig. S2.** *(A)* 2θ-ω x-ray scan of 30 uc (12 nm) SrIrO₃ thin film on SrTiO₃(001) substrate showing single-phase SrIrO₃ with thickness oscillations indicating a smooth surface and sharp interface with the substrate. *(B)* Φ-scan of the SrIrO₃ film showing its equivalent pseudocubic epitaxial arrangement with the underlying SrTiO₃ substrate. *(C)* reciprocal space mapping around the (1̄03) SrTiO₃ substrate peak confirming the fully coherent in-plane lattice of the SrIrO₃ film on the SrTiO₃ substrate.

## II. Crystallographic domain structure of SrIrO₃ on SrTiO₃ (001) substrate

It is known(1) that orthorhombic perovskites like SrIrO₃ with *Pbnm* space group symmetry orient themselves on cubic substrates with [110]ₒ out of plane along [001] with [11̄0]ₒ and [001]ₒ in-plane along [100] and [010], respectively, as depicted in Fig. S3A. Such an epitaxial arrangement produces a distortion of the orthorhombic unit cell due to the compressive/tensile strain along [11̄0]ₒ, which causes the orthorhombic film to assume a slightly distorted monoclinic structure with $\alpha = \beta = 90° \neq \gamma$. In the case of SrIrO₃ films on SrTiO₃, the pseudocubic lattice parameter of SrIrO₃ is ~3.95 Å compared to 3.905 Å for SrTiO₃, which leads to an in-plane compressively strained SrIrO₃ film. This compressive strain reduces γ below 90°. This monoclinic distortion of the orthorhombic unit cell produces an out-of-plane tilt in the thin film relative to the substrate due to the octahedral tilt in the thin film, which can be measured from reciprocal space mappings, as described next(1). It has been shown that for the case of SrRuO₃ thin films grown on SrTiO₃ (001) substrates that a symmetry change from orthorhombic to tetragonal occurs with decreasing film thickness due to a significant partial suppression of the SrRuO₃ octahedral tilt in the thinner film

limit(2). Since SrIrO$_3$ is structurally and chemically similar to SrRuO$_3$, a similar change in symmetry was expected.

By examining the {103}$_{pc}$ film and substrate reflections at $\Phi$=90° increments, it is thus possible to determine this relative tilt in the orthorhombic films by comparing the peak position in $L$ at each $\Phi$ angle, as their film peak positions will show deviations in $Q_x$, the surface-normal component of the x-ray scattering vector. This will shift the peak position in $L$ along 2 of the $\Phi$-angle peaks, whereas the other 2 alignments will have identical peak positions in $L$. The $L$-shifts will exist along the [103]$_{pc}$ and [$\bar{1}$03]$_{pc}$ film peaks since the strain that creates the distorted tilt lies in-plane along [100]$_{pc}$. Therefore, the (013)$_{pc}$ and (0$\bar{1}$3)$_{pc}$ reflections should exhibit the same film peak position in $L$, since no tilt exists along this direction. As can be seen in Fig. S3B, the splitting from (103)$_{pc}$ and ($\bar{1}$03)$_{pc}$ is pronounced at 20 uc, but is slowly suppressed as the film thickness decreases. At 8 uc, the {103}$_{pc}$ family shows no deviation in $L$, indicating that relatively no tilt exists, which means that a global tetragonal symmetry is established by significant suppression of the octahedral tilt. From the position of these $L$ peaks, geometrical analysis was performed to calculate the $a/b$ lattice parameter ratio(1) presented in Fig. 4a of the main text, as the relative lengths of a and b arise from the tilt.

However, while such scans are effective for determining the tilt from epitaxial strain, they ignore tilting of the octahedra from orthorhombic domains. Since lower symmetry perovskites like orthorhombic SrIrO$_3$ show tilts and rotation patterns in their octahedra along all 3 pseudocubic directions(3), they will exhibit extra x-ray reflections(4) between pseudocubic peaks that arise from doubling the unit cell along particular crystallographic directions. Therefore, based on the bulk tilt pattern of SrIrO$_3$, scans to look for the (221)$_o$ reflection were performed to check if the films were completely tetragonal with no octahedral tilt. This orthorhombic reflection corresponds to a half-order {$^1/_2$02}$_{pc}$ family of reflections that do not exist in non-tilted perovskite systems. Thus, any measured intensity from the (221)$_o$ peak would indicate the presence of orthorhombic domains in the films. In Fig. S3C we see that intensity from the measured (221)$_o$ reflection in our SrIrO$_3$ films persists even in the 8 uc film that showed no tilt from the {103}$_{pc}$ reciprocal space mappings. It should be noted, however, that although the (221)$_o$ peak is still observable in the thin 8 uc film, the intensity of this peak drops much more quickly with decreasing thickness than the primary (103)$_{pc}$ peak from 20 to 8 uc. If this (221)$_o$ intensity change were solely due to the progressively thinner SrIrO$_3$ films, the (103) should decrease at the same rate, which is not the case, as shown in Fig. S3D. Thus, this signifies a global partial suppression of the octahedral tilts and rotations. Thus, while the 8 uc film may retain small octahedral tilts (this was also verified from RHEED experiments on the SrIrO$_3$ surface along the (1,0) and (0,1) pseudocubic directions), these domains are greatly suppressed as the global structure clearly tends towards the tetragonal symmetry from 20 to 8 uc. It should be mentioned, lastly, that the reflection of this {221}$_o$ family was observed at 90° increments of $\Phi$, which correspond to {$^1/_2$02}$_{pc}$ and {0$^1/_2$2}$_{pc}$. This finding illustrates that a mixture of [1$\bar{1}$0]$_o$ and [001]$_o$ SrIrO$_3$ domains lie along the substrate [100] and [010] directions, which was also confirmed by RHEED. If the film were truly single-crystal, these half-order reflections would only occur at 180° increments(4). However, the domain preference in the 20 uc film (which showed the highest spin-torque ratio) was ~85% along [1$\bar{1}$0]$_o$ ([100]) direction, indicating that the SrIrO$_3$ film was close to single-crystal.

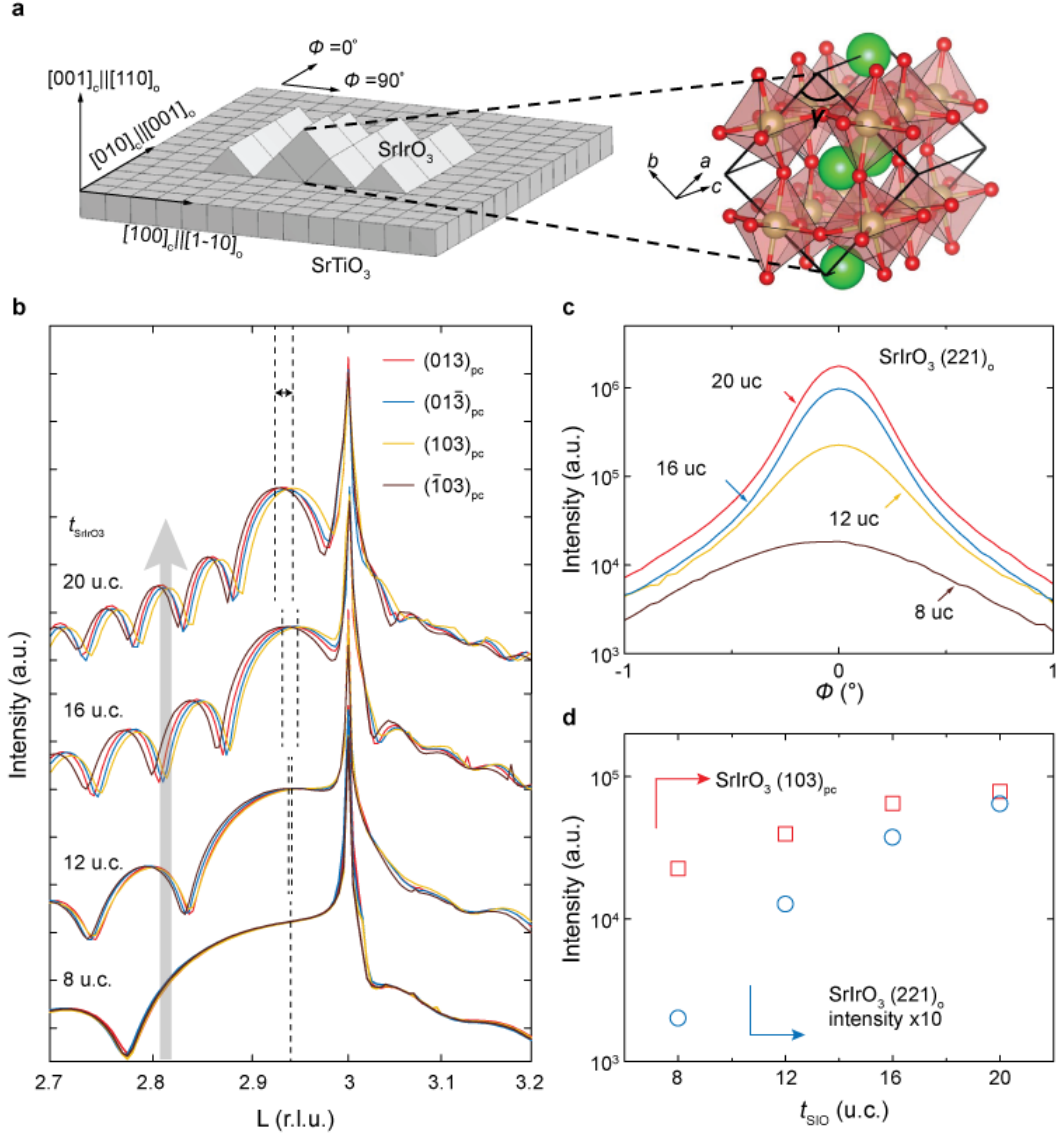

**Fig. S3.** *(A)* Epitaxial arrangement of orthorhombic SrIrO$_3$ on cubic SrTiO$_3$ with SrIrO$_3$ [1-10]$_o$, [001]$_o$, and [001]$_o$ along SrTiO$_3$ [100], [100], and [001], respectively. *(B)* Integrated reciprocal space mappings around the {103}$_{pc}$ pseudocubic reflection of 8, 12, 16, and 20 uc SrIrO$_3$ films on SrTiO$_3$ (001) substrates. The separation of the (103)$_{pc}$ and ($\bar{1}$03)$_{pc}$ film peak positions, which indicates an out-of-plane tilt of the SrIrO$_3$ film due to the epitaxial strain from the substrate, is shown to decrease to zero from 20 uc to 8 uc, indicating a partial suppression in SrIrO$_3$ of the distorted orthorhombic tilt to a near tetragonal-like structure. *(C)* Intensity of orthorhombic (221)$_o$ SrIrO$_3$ peak, which decreases from 20 to 8 uc. *(D)* Intensity comparison of the SrIrO$_3$ (103) pseudocubic and (221)$_o$ orthorhombic peak intensities as a function of film thickness.

### III. ST-FMR line shape

The ST-FMR signal with the current-induced in-plane and out-of-plane torque components can be described by the Landau–Lifshitz–Gilbert–Slonczewski equation(5),

$$\frac{d\vec{m}}{dt} = -\gamma \vec{m} \times \mu_0 \vec{H}_{eff} + \alpha \vec{m} \times \frac{d\vec{m}}{dt} + \tau_\perp \vec{\sigma} \times \vec{m} + \tau_\parallel \vec{m} \times (\vec{\sigma} \times \vec{m}) \qquad [S1]$$

where $\gamma$ is the gyromagnetic ratio, $\mu_0$ is the permeability in vacuum, $\vec{H}_{eff}$ is the effective magnetic field including the external magnetic field $\vec{H}_{ext}$ and the demagnetization field, $\alpha$ is the Gilbert damping coefficient, and $\tau_\perp$ and $\tau_\parallel$ are the out-of-plane and in-plane torque components as illustrated in Fig. 2A. The ST-FMR mixing voltage can be then written in the form as,

$$V_{mix} = S \frac{W^2}{(\mu_0 H_{ext} - \mu_0 H_{FMR})^2 + W^2} + A \frac{W(\mu_0 H_{ext} - \mu_0 H_{FMR})}{(\mu_0 H_{ext} - \mu_0 H_{FMR})^2 + W^2} \qquad [S2]$$

where $W$ is the half-width-at-half-maximum resonance linewidth, and $H_{FMR}$ is the resonance field. $S$ and $A$ are the symmetric and antisymmetric amplitude of the Lorentzian, and can be expressed as,

$$S = -\frac{I_{rf}}{2} \left(\frac{dR}{d\varphi}\right) \frac{1}{\alpha(2\mu_0 H_{FMR} + \mu_0 M_{eff})} \tau_\parallel \qquad [S3]$$

$$A = -\frac{I_{rf}}{2} \left(\frac{dR}{d\varphi}\right) \frac{\sqrt{1 + M_{eff}/H_{FMR}}}{\alpha(2\mu_0 H_{FMR} + \mu_0 M_{eff})} \tau_\perp, \qquad [S4]$$

where $I_{rf}$ is the microwave current, $R(\varphi)$ is the device resistance as a function of in-plane magnetic field angle $\varphi$ due to the anisotropic magnetoresistance of Py, and $\mu_0 M_{eff}$ is the effective magnetization. Fig. S4A shows a typical ST-FMR spectrum for a 3.5 nm Py/8 nm (20 uc) SrIrO$_3$ sample. The resonance line shape is well fitted to a sum of symmetric and antisymmetric Lorentzian components (red and yellow dashed curves). The magnetic field angle $\varphi$ dependence of $S$ and $A$ amplitudes are shown in Fig. S4B, where the symmetric and antisymmetric components both depend on $\varphi$ according to the form $\sin(2\varphi)\cos\varphi$. This can be interpreted as the product of the contributions from the AMR in Py $[dR/d\varphi \propto \sin(2\varphi)]$ and the current-induced torque ($\tau \propto \cos$). Then the magnitude of torque components can be determined by extracting the symmetric and antisymmetric amplitude from Eq. S3 and Eq. S4. The two torque ratios can be calculated as $\theta_{\parallel/\perp} = \tau_{\parallel/\perp} M_s t \frac{I_{rf} R}{l\cos(\varphi)} \left(\frac{2e}{\hbar}\right) \rho$, where $M_s$ and $t$ are the saturation magnetization and the thickness of Py; $l$ is the length of the device bar, $\hbar$ is the reduced Planck's constant, $e$ is the electron charge. However, due to the SrTiO$_3$ substrate microwave current shunting issue (described in Methods), the spin-torque ratio obtained by the ST-FMR lineshape analysis may not be reliable.

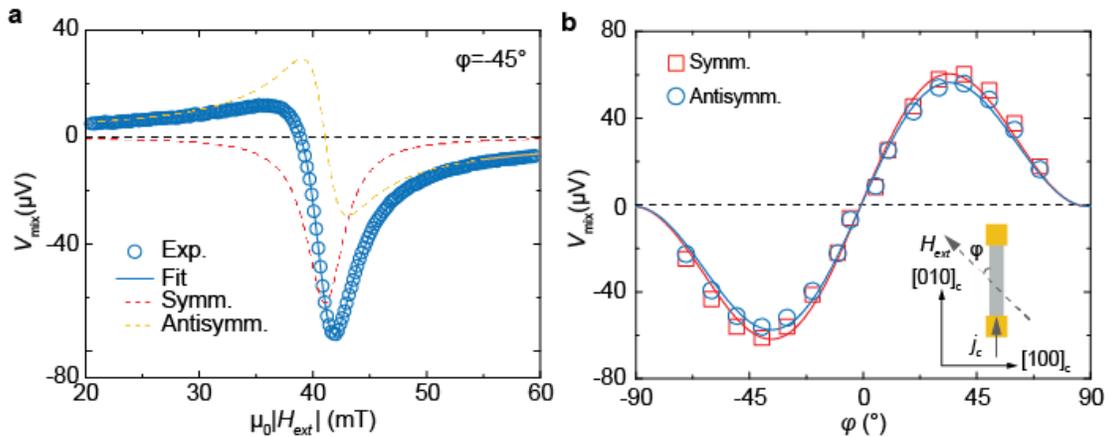

**Fig. S4.** *(A)* ST-FMR spectrum (fitted to a Lorentzian function, solid line) for a 3.5 nm Py/8 nm (20 uc) SrIrO₃ sample (20 μm × 40 μm) with microwave current applied along the [1$\bar{1}$0]ₒ axis. The resonance spectrum is obtained at a fixed microwave frequency, and with an in-plane external magnetic field swept through the ferromagnetic resonance condition in Py. The dashed lines represent the fits of the symmetric and antisymmetric components. The external magnetic field is oriented at an angle *φ*=-45° with respect to the current axis. The applied microwave frequency and power are 5.5 GHz and 12 dBm, respectively. The $V_{mix}$ across the device bar is acquired by a dc voltage meter. *(B)* Symmetric and antisymmetric resonance components as a function of the external magnetic field angle φ, which are fitted to sin(2φ)cos(φ).

## IV. MOKE measurements

The out-of-plane component of the magnetization $m_z$ of an in-plane magnetized thin film magnet along its width under the influence of a current-induced anti-damping torque and the Oersted field also arising from that same current is given by

$$m_z(x) = \frac{\hbar}{2eM_st_{FM}}\frac{J_cX\theta_{||}cos\varphi}{B_{ext}+M_{eff}} - \frac{\mu_0J_ct_{SIO}}{4\pi(B_{ext}+M_{eff})}\left(ln\left(t_{tot}^2+\left(x+\frac{w}{2}\right)^2\right)-ln\left(t_{tot}^2+\left(x-\frac{w}{2}\right)^2\right)\right) \quad [S5]$$

, where $J_c$ is the total current density through the sample assuming uniform current distribution, $X$ is the fraction of the total current which flows through spin Hall material (determined as described in the following section), $φ$ is the angle of the magnetization with respect to the current flow direction, $t_{tot}$ is the total thickness of all layers, $x$ is the position across the width of the device (perpendicular to the current flow direction) with its origin at the device midpoint, and $w$ is the width of the device(6). We measure $m_z$ via polar MOKE by locking into the oscillation of the Kerr rotation as a result of applying a low frequency (5667 Hz) excitation current to the sample resulting in an RMS current of 7-10 mA. Because can calculate the $m_z$ expected for the Oersted field *a priori*, we can use the amplitude of the Oersted response to "self-calibrate" our analysis making it robust to variations in magneto-optical response from sample to sample.

We do two scans across the device (perpendicular to the current flow direction) with all parameters equal except with opposite signs of the external field. The sum of the signal of these two traces then only contains information on the Oersted field contribution (second term in the above equation, even in field) while the difference of these two traces contains information on the anti-damping torque only (odd in field). Fig. S5 shows the cumulative integral signal across the width of the device for a 5.5 nm Py/10 nm (25 uc) SrIrO₃ sample. Fitting to the above equation numerically convolved with a Gaussian beam profile is perilous and contraindicated owing to the low signal to noise. To surmount these difficulties we note that the thickness of the device is much less than width. In this case, if we take the absolute value of the area underneath each trace, then the ratio of the areas gives $\theta_{||}=\frac{A_{diff}}{A_{sum}}\frac{e}{\hbar}\frac{\mu_0M_st_{FM}t_{tot}ln(4)}{X\pi}$. We validate this analysis on a 6nm Py/ 8 nm Pt control sample and find a $\theta_{||}$ of 0.065±0.01, consistent with earlier results using other techniques for this system(7).

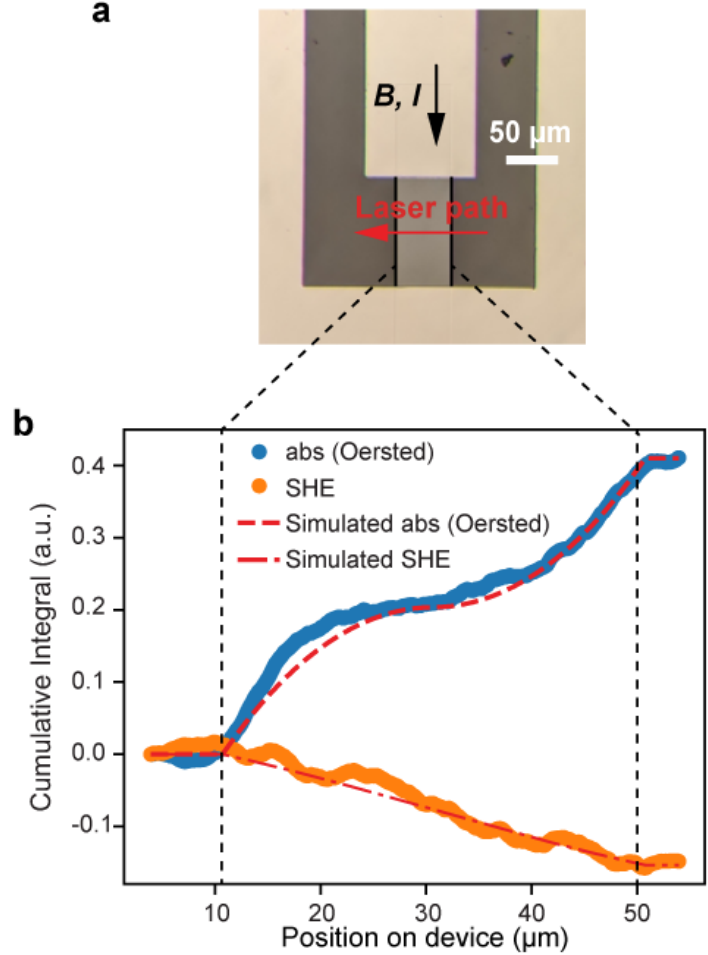

**Fig. S5.** *(A)* Optical images of the sample geometry for the MOKE measurement. The black arrow shows the magnetic field and the current direction, and the red arrow shows the laser path. *(B)* Cumulative integrals of the MOKE response of a representative scan as a function of the ~10 um laser spot position across the device. The Oersted signal (blue) derives from the out of plane component of the magnetic field generated by the applied current and is used to "self-calibrate" the signal due to the spin Hall effect (orange) which corresponds to a spatially uniform out-of-plane effective magnetic field (or an in-plane torque) as described in the text. The dashed lines are simulated cumulative integrals using the spin Hall efficiency derived from the ratio of the total integrals as described in the text and the applied RMS current of 7 mA.

## V. SrIrO₃ thin films electrical transport property

The bare SrIrO₃ thin film transport property was measured by using the van der Pauw technique in 5 mm by 5 mm SrIrO₃//SrTiO₃ samples. The SrIrO₃ room temperature resistivity shows a slight sample-to-sample variation. To determine the anisotropy of the transport property of SrIrO₃, the sheet resistance of the SrIrO₃ thin film was measured by using a 4-point resistance technique on Hall bars patterned along the [100]$_{pc}$ and [010]$_{pc}$ axes. Typical SrIrO₃ resistivity versus temperature curves are shown in Fig. S6A exhibiting metallic transport characteristics in both crystalline orientations.

The Py resistivity was measured by using the van der Pauw technique in Al₂O₃/Py//SrTiO₃ reference samples. The resistivity for the 3.5 nm Py in our work is 62 μΩ cm. To determine the current fraction of SrIrO₃ in each Py/SrIrO₃ bilayer sample, we treat the Py and SrIrO₃ layers as parallel resistors. The SrIrO₃ resistivity and its current fraction are estimated by assuming that the Py resistivity is constant among different samples. Fig. S6B shows the estimated SrIrO₃ resistivity and its current fraction as a function of SrIrO₃ thickness t_{SIO} in 3.5 nm Py/ t_{SIO} SrIrO₃//SrTiO₃ samples.

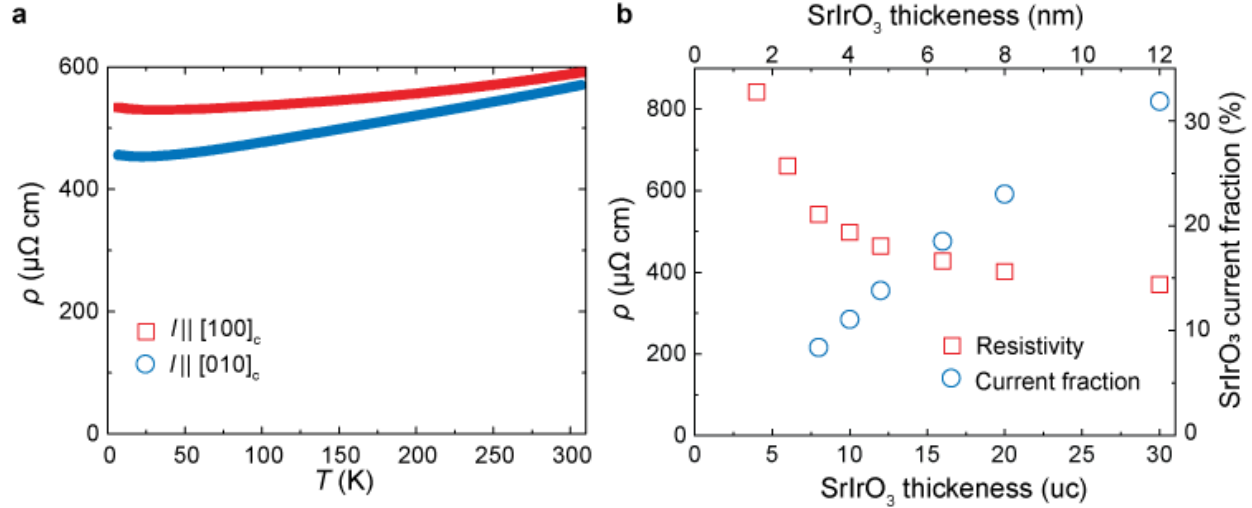

**Fig. S6.** *(A)* Temperature dependence of the resistivity of the 20 uc SrIrO₃//SrTiO₃ sample with the current applied along both SrTiO₃ [100] and [010] directions. *(B)* SrIrO₃ room temperature resistivity and the current fraction of the 3.5 nm Py/ SrIrO₃//SrTiO₃ with various SrIrO₃ thicknesses.